\documentclass{ws-procs9x6}            
\begin{document}
\title{Light meson spectroscopy at $e^+e^-$ machines}

\author{B. J. Liu$^*$}

\address{Institute of High Energy Physics, Chinese Academy of Sciences\\
Beijing, 100049, China\\
$^*$E-mail: liubj@ihep.ac.cn\\
}

\begin{abstract}
The study of light hadrons is central to the understanding of confinement--a unique property of QCD. The quark model describs mesons as bound states of quarks and antiquarks. LQCD and QCD-motivated models for hadrons, however, predict a richer spectrum of mesons that takes into account not only the quark degrees of freedom but also the gluonic degrees of freedom. A selection of recent progress in the light-quark sector with unprecedented high-statistics data sets from $e^+e^-$ experiments are reviewed.
\end{abstract}

\keywords{Light QCD exotics; Glueball; $e^+e^-$ experiments.}

\bodymatter

\section{Introduction}
Confinement is a unique property of QCD.  The spectrum of light hadrons serves as an excellent probe of QCD in the confinement regime ~\cite{Brambilla:2014jmp,Meyer:2010ku,Crede:2008vw,Klempt:2007cp,Amsler:2004ps,Godfrey:1998pd}. The quark model describs mesons as bound states of quarks and antiquarks. LQCD and QCD-motivated models for hadrons, however, predict a richer spectrum of mesons that takes into account not only the quark degrees of freedom but also the gluonic degrees of freedom. Light mesons can be produced at $e^+e^-$ experiments through $e^+e^-$ annihilation, through the use of the radiative return method and through two-photon fusion. Unprecedented high-statistics data sets from experiments at $e^+e^-$ machines provide great opportunities to the quantitative understanding of confinement.
\section{Search for glueballs}
The spectrum of glueballs both from the quenched lattice QCD studies~\cite{Morningstar:1999rf, Chen:2005mg}
and the full-QCD study~\cite{Gregory:2012hu, Sun:2017ipk} show that the lightest one having scalar quantum
numbers $0^{++}$ and a mass between 1.5~GeV and 1.7~GeV.
Also the next-higher glueball
states have nonexotic quantum numbers,
$2^{++}$ (mass 2.3--2.4~GeV)
and $0^{-+}$ (mass 2.3--2.6~GeV), and hence will be mixed into the conventional meson spectrum and difficult to be identified experimentally.
It requires systematic studies  to identify a glueball by
searching for outnumbering of conventional quark model states and comparing a candidate¡¯s properties to the expected
properties of glueballs and conventional mesons.
Glueballs are expected to appear in so-called gluon-rich environments. The radiative decays of the $J/\psi$ meson provide such a gluon-rich environment and are therefore regarded as one of the most
promising hunting grounds for glueballs. Recent LQCD calculations predict that the partial width of $J/\psi$ radiatively decaying into the pure gauge scalar glueball is 0.35(8) keV, which corresponds to a branching ratio of $3.8(9)\times 10^{-3}$~\cite{Gui:2012gx}; the partial decay width for a tensor glueball is estimated to be 1.01(22)(10) keV which corresponds to a large branching ratio $1.1(2)(1)\times10^{-2}$~\cite{Yang:2013xba}.
Even though the fact of a supernumerary  state
is suggestive for the mixing of glueball with $q\bar{q}$ state, the decay rates and production mechanisms are also needed to unravel the quark content of $f_0(1500)$ and $f_0(1710)$.
In the PWA of $J/\psi\to\gamma\eta\eta$~\cite{Ablikim:2013hq} and $J/\psi\to\gamma K_S K_S$~\cite{Ablikim:2018izx} at BESIII, the branching fractions of the $f_0(1710)$ are one order of magnitude larger than those of the $f_0(1500)$. With the new measurements from BESIII, the known branching fraction of $J/\psi\to\gamma f_0(1710)$ ~\cite{PDG} is up to ($1.7\times 10^{-3}$, which is already comparable to the LQCD calculation of scalar glueball ($3.8(9)\times 10^{-3}$~\cite{Gui:2012gx}). The production property suggests $f_0(1710)$ has large gluonic component than $f_0(1500)$. 
The two-photon width can be used to identify glueball, even though it can also be adjusted by the glueball mixing with $q\bar{q}$ state ~\cite{Cheng:2015iaa}. The scalar meson $f_0(1710)$ has
been seen in $\gamma\gamma\to K_S K_S$ ~\cite{Acciarri:2000ex,Uehara:2013mbo}. $f_0(1500)$ was not seen in $\gamma\gamma\to K_S K_S$
by L3 ~\cite{Acciarri:2000ex}, or in $\gamma\gamma\to\pi^+\pi^-$, by ALEPH ~\cite{Barate:1999ze}. However,
a resonance observed in $\gamma\gamma\to\pi^0\pi^0$ by Belle ~\cite{Uehara:2008ep} is close
to the $f_0(1500)$ mass, though it is also consistent with
$f_0(1370)$ because of the large errors in the experiment and
the large uncertainty in the $f_0(1370)$ mass. The assignment of scalars in two-photon fusion requires further studies with better precision. It is noticeable the nonobservation of $f_0(1710)$
and observation of $f_0(1500)$ in $B_s\to J/\psi \pi \pi$ by LHCb~\cite{Aaij:2019mhf}. Because of the spectator s quark of $B_s$, the isosinglet scalar resonance $f_0$ produced in $B_s\to J/\psi f_0$ decays should have a sizable $s\bar{s}$ component~\cite{Lu:2013jj}.
\begin{figure*}[t!]
\includegraphics[width=\textwidth]{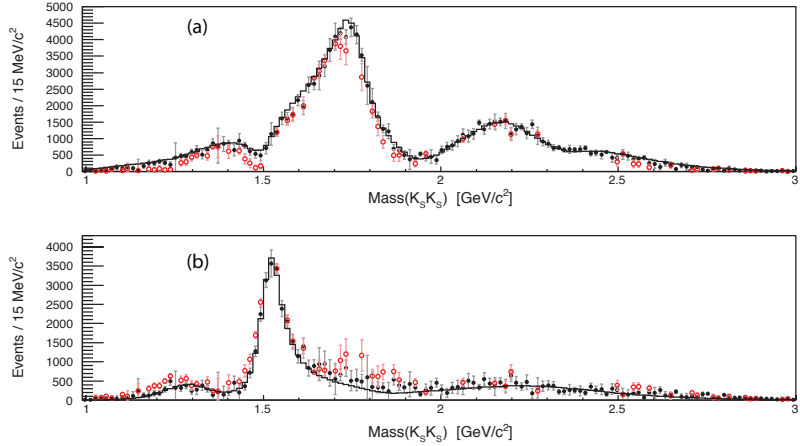}
\caption{\label{fig:inten_comp} (Color online) Intensities for the total (a) $0^{++}$ and (b) $2^{++}$
amplitudes in $J/\psi\rightarrow \gamma K_S K_S$ from BESIII as a function of $K_{S}K_{S}$ invariant mass for the nominal results without acceptance correction~\cite{Ablikim:2018izx}. The solid black markers show
one set of solutions from the mass-independent analysis, while the open red markers represent its ambiguous
partner and the histogram shows the results of the mass-dependent analysis.}
\end{figure*}
The three tensors $f_2(2010)$, $f_2(2300)$ and $f_2(2340)$ observed
in $\pi^- p\rightarrow \phi\phi n$~\cite{bibpiN} are also observed
in $J/\psi\rightarrow\gamma\phi\phi$~\cite{Ablikim:2016hlu}.
 The large production rate of the $f_2(2340)$
in $J/\psi\rightarrow\gamma\phi\phi$ and $J/\psi\rightarrow\gamma\eta\eta$~\cite{Ablikim:2013hq} indicates $f_2(2340)$ is a good candidate of tensor glueball.  Significant tensor contribution around 2.4~GeV also presents in $J/\psi\to\gamma\pi^0\pi^0$~\cite{Ablikim:2015umt} and  $J/\psi\to\gamma K_S K_S$~\cite{Ablikim:2018izx}. However, the measured production rate of $f_2(2370)$ appears
to be substantially lower than the LQCD calculated
value~\cite{Yang:2013xba}. It is desirable to search for more decay modes  to establish
and characterize the lowest tensor glueball.
\section{$a_0(980)-f_0(980)$ mixing}
\begin{figure}[tbp]
  \begin{center}
    \includegraphics[width=0.5\columnwidth]{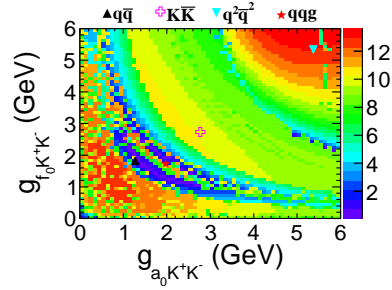}
    \caption{\label{signif}The statistical significance of the signal scanned in the two-dimensional space of $g_{a_{0}K^{+}K^{-}}$ and $g_{f_{0}K^{+}K^{-}}$. The regions with higher statistical significance indicate larger probability for the emergence of the two coupling constants. The markers indicate predictions from various illustrative theoretical models.}
  \end{center}
\end{figure}

After the discoveries of $a_0(980)$ and $f_0(980)$ several decades ago, explanations about the nature of these two light scalar mesons have still been controversial. These two states, with similar masses but different decay modes and isospins, are difficult to accommodate in the traditional quark-antiquark model, and many alternative formulations have been proposed to explain their internal structure, including tetra-quarks~\cite{Jaffe:1976ig,Alford:2000mm,Maiani:2004uc,Maiani:2007iw,Hooft:2008we}, $K\bar{K}$ molecule~\cite{Weinstein:1990gu}, or quark-antiquark gluon hybrid~\cite{gluon}. Further insights
into $a_0(980)$ and $f_0(980)$ are expected from their mixing~\cite{Achasov:1979xc}.  The mixing mechanism in the system of $a_0(980)-f_0(980)$ is considered to be a sensitive probe to clarify the nature of these two mesons. In
particular, the leading contribution to the isospin-violating mixing
transition amplitudes for $f_{0}(980)\to a^{0}_{0}(980)$ and
$a^{0}_{0}(980)\to f_{0}(980)$, is shown to be dominated by the
difference of the unitarity cut which arises from the mass difference
between the charged and neutral $K\bar{K}$ pairs. As a consequence, a
narrow peak of about 8 MeV$/c^2$ is predicted between the charged and
neutral $K\bar{K}$
thresholds.
The corresponding signal is predicted
in the isospin-violating processes of $J/\psi\to\phi a^{0}_{0}(980)$~\cite{Wu:2007jh,Hanhart:2007bd} and $\chi_{c1}\to\pi^{0} f_{0}(980)$~\cite{Wu:2008hx}.
The signals of $f_0(980)\to a^0_0(980)$ and $a^0_0(980)\to f_0(980)$ mixing are first observed in $J/\psi\to\phi f_{0}(980)\to\phi a^{0}_{0}(980)\to\phi\eta\pi^{0}$ and $\chi_{c1}\to\pi^{0} a^{0}_{0}(980)\to\pi^{0} f_{0}(980)\to\pi^{0}\pi^{+}\pi^{-}$ at BESIII ~\cite{Ablikim:2010aa, Ablikim:2018pik}. 
The statistical significance of the signal versus the values of $g_{a_{0}K^{+}K^{-}}$ and $g_{f_{0}K^{+}K^{-}}$ is shown in Fig.~\ref{signif}. The regions with higher statistical significance indicate larger probability for the emergence of the two coupling constants. This direct measurement of $a_0(980)-f_0(980)$ mixing is a sensitive probe to the internal structure of those ground state scalars and sheds important light on their nature. The new results from BESIII provide critical constraints to the development of theoretical models for $a_0(980)$ and $f_0(980)$. It is theorists¡¯ turn to refine the calculations to understand the inner structure of the $a_0(980)$ and $f_0(980)$ mesons.
\section{$X(1835)$ , $X(p\bar{p})$ and structures near $N\bar{N}$ threshold in $e^+e^-$ cross sections}
\begin{figure}[htbp]
  \centering
    \includegraphics[width=0.48\textwidth]{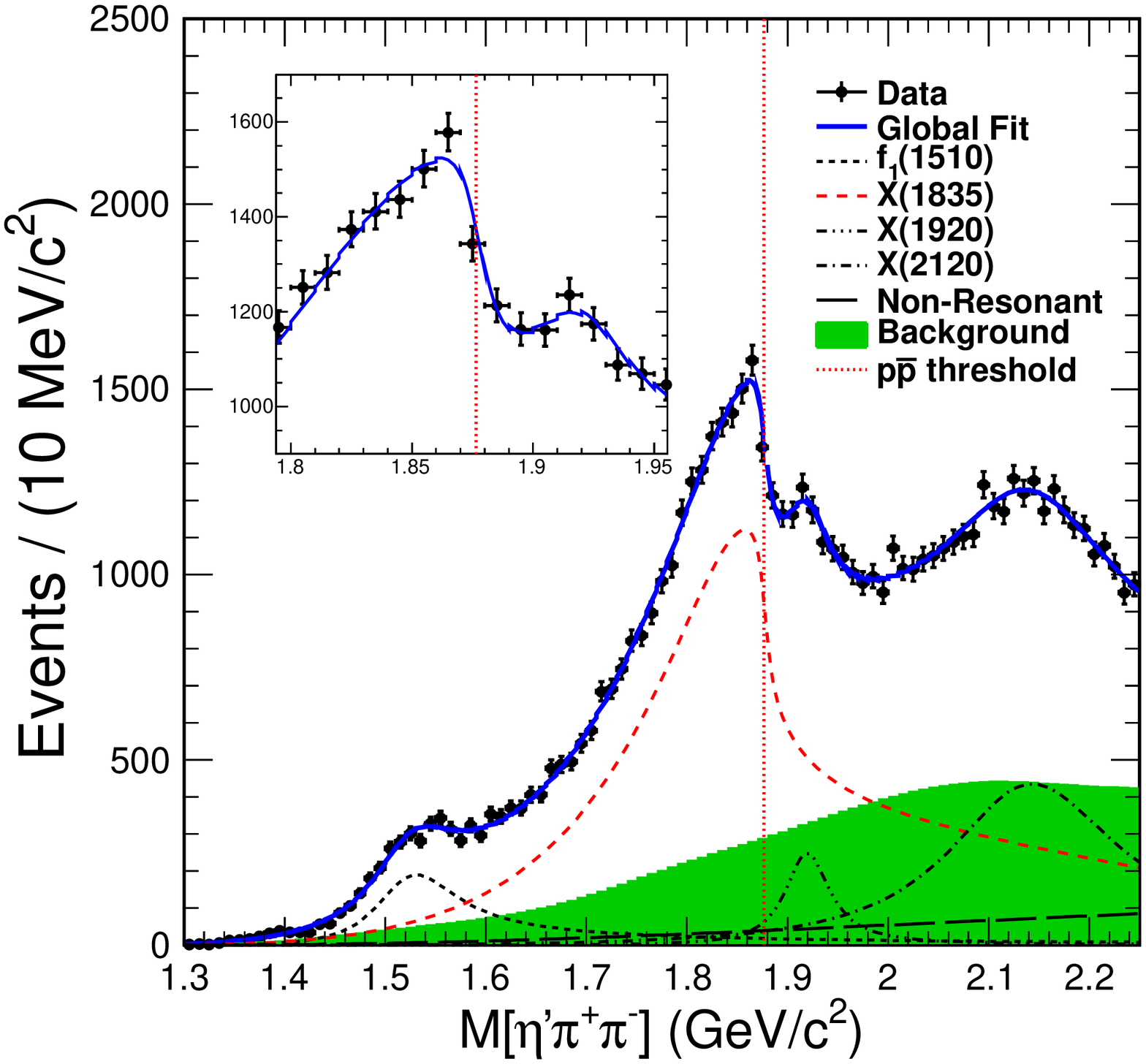}
  \includegraphics[width=0.48\textwidth]{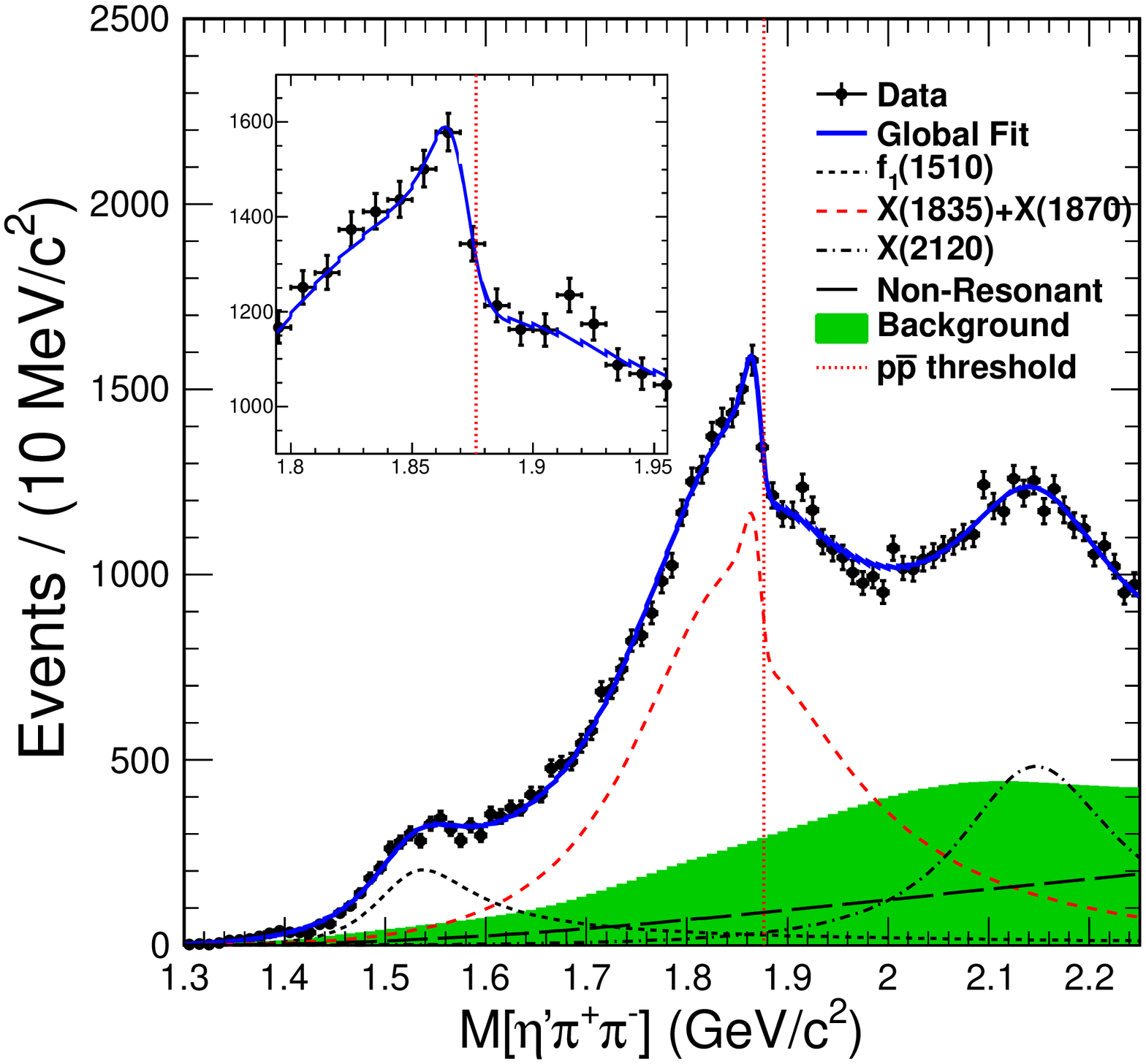}
  \caption{An anomalous line shape of the $\eta^{\prime}\pi^{+}\pi^{-}$ mass spectrum near the $p\bar{p}$ mass threshold in $J/\psi\rightarrow\gamma\eta^{\prime}\pi^{+}\pi^{-}$. (a) shows
  the fitting results with Flatt\'{e} formula and (b) shows the fitting results with the coherent sum of two Breit-Wigner amplitudes }
  \label{x1835}
\end{figure}

An anomalously strong enhancement at the proton-antiproton ($p\bar{p}$)
mass threshold, dubbed $X(p\bar{p})$,
was first observed by BES in $J/\psi\rightarrow\gamma p\bar{p}$ decays~\cite{xpp_bes2};
this observation was confirmed by BESIII~\cite{xpp_bes3} and CLEO~\cite{xpp_cleo}, while no evidence has been seen in other channels, such as $J/\psi\to\omega p\bar{p}$~\cite{Ablikim:2007ac,Ablikim:2013cif}
 or $J/\psi\to\phi p\bar{p}$~\cite{Ablikim:2015pkc}. These non-observations disfavor the mass-threshold enhancement
attribution to the effects of $p\bar{p}$ final state interactions.
This enhancement was subsequently determined to have spin-parity
$J^{P}=0^{-}$ by BESIII~\cite{xpp_bes3pwa}.
The state $X(1835)$ was first observed by the BES experiment as a peak
in $J/\psi\rightarrow\gamma\eta^{\prime}\pi^{+}\pi^{-}$ decays~\cite{x1835_bes2}. This observation was later
confirmed by BESIII~\cite{x1835_bes3} and was also observed in the $\eta K^{0}_{S} K^{0}_{S}$ channel, where its spin-parity was determined to be $J^{P}=0^{-}$ by a partial
wave analysis~\cite{x1835_qiny}. No evidence of $X(1835)$ is found in $J/\psi\to\omega\eta^{\prime}\pi^{+}\pi^{-}$ ~\cite{BESIII:2019sfz}.
$\eta(1475)\to\gamma\phi$ and $X(1835)\to\gamma\phi$ are observed in the decay of $J/\psi\to\gamma\gamma\phi$ at BESIII ~\cite{Ablikim:2018hxj}, which indicates that both $\eta(1475)$ and $X(1835)$ contain a sizeable $s\bar{s}$ component.
Using high-statistics $J/\psi$ events, BESIII studied the $J/\psi\to\gamma\eta^{\prime}\pi^{+}\pi^{-}$ process and
  observed a significant abrupt change in the slope of the $\eta^{\prime}\pi^{+}\pi^{-}$ invariant mass distribution at the
  proton-antiproton ($p\bar{p}$) mass threshold~\cite{Ablikim:2016itz}.
  Two models are used to characterize the $\eta^{\prime}\pi^{+}\pi^{-}$ line shape
  around 1.85~GeV$/c^{2}$: one which explicitly incorporates the opening of a
  decay threshold in the mass spectrum (Flatt\'{e} formula) (Fig.~\ref{x1835}(a)), and another
  which is the coherent sum of two resonant amplitudes(Fig.~\ref{x1835} (b)).
  Both fits show almost equally good agreement with data,
  and suggest the existence of either a broad state with strong couplings to $p\bar{p}$ final states
  or a narrow state just below the $p\bar{p}$ mass threshold.
Without imposing a relation to the $0^{-+}$ structures near $N\bar{N}$ threshold observed in $J/\psi$ radiative decays, we should notice the dip near $p\bar{p}$ mass threshold in the six-pion photon production cross section observed by DM2~\cite{Bisello:1981sh}, Focus~\cite{Frabetti:2003pw}, and BaBar~\cite{Aubert:2006jq}, which is $1^{--}$ . Recently, a fine structure with about 1 MeV width at the nucleon-antinucleon threshold in $e^+e^- \to 3(\pi^+\pi^-)$ and $e^+e^- \to K^+ K^-\pi^+\pi^-$ production is observed by CMD-3~\cite{CMD-3:2018kql}, while no such structure is seen in $e^+e^- \to 2(\pi^+\pi^-)$.  However, cross sections of $p\bar{p}$ annihilation into $2(\pi^+\pi^-) > 3(\pi^+\pi^-) \gg K^+ K^-\pi^+\pi^-$, which suggests  a more complicated dynamics in  $e^+e^- \to hadrons $ at the nucleon-antinucleon threshold.
\section{Outlook}
Data from $e^+e^-$  machines with unprecedented statistical accuracy provides great opportunities to map out light mesons as complete and as precise as possible.
In the next few years, many experiments (BESIII, CMD-3, SND,~{\it etc.}) will  continue to be active,  while  BelleII has already started data taking. BESIII collected 10 billions of $J/\psi$, which is unique for studying and searching for QCD exotics as a gluon rich environment with clearly defined initial and final state properties. We can expect a continuous flow of interesting results and new insights into QCD in the confinement regime from the meson spectroscopy at $e^+e^-$ machines.

\end{document}